\documentclass[epj,referee]{svjour}
\usepackage{graphics}
\begin{document}

\title{Spin glass models with Kac interactions}

\author{Silvio Franz}

\institute{  Laboratoire de Physique Th\'eorique et Mod\`eles Statistiques,
b\^at. 100, Université Paris-Sud 11,
Centre scientifique d'Orsay,
15 rue G. Cl\'emenceau,
91405 Orsay cedex France
}

\date{\today}

\abstract{
  In this paper I will review my work on disordered systems -spin
  glass model with two body and $p>2$ body interactions- with long but
  finite interaction range $R$. I will describe the relation of these
  model with Mean Field Theory in the Kac limit and some attempts to
  go beyond mean field. 
}


\maketitle

\section{Introduction}
In statistical physics mean-field theories play the important role of
offering a first rough approximate scheme to understand collective
phenomena and phase transitions. However several well known
pathologies plague this description, which can be traced in the fact
that the finite range character of the interactions of physical
systems is neglected. Many of the progresses in the comprehension of
the physics of pure (i.e. non-disordered) equilibrium and dynamical
systems in the statistical physics of 20th century can be viewed as
amendment and extensions of mean-field schemes to account of more
complex behavior. 
A remarkable example is given by nucleation theory
of first order transitions and phase coexistence that can be viewed as
a non-perturbative expansion in the inverse of the interaction range.
The role of an infinite interaction range in the mean-field
description of first-order transition and metastability phenomena, had
been emphasized in classical papers by Kac, Uhlenbeck and Hemmer
\cite{kuh} in 1D and Lebowitz and Penrose \cite{lp} in arbitrary D,
showing that when range of interaction $R$ is sent to infinity {\it
  after} the thermodynamic limit, mean-field description emerges,
complemented by the Maxwell construction that eliminates possible unphysical
thermodynamical instability of mean-field theory.

In this contribution I will discuss the relation between mean-field
theory and finite D physics in spin glass models and generalized spin
glass models with $p$-body interactions.  Mean field theory for these
systems predicts ergodicity breaking with the coexistence of many
phases unrelated by physical symmetry.  However for finite range spin
glasses, the mere existence of an upper critical dimension above which
there is a low temperature spin glass phase with the MF
characteristics has been questioned and is still matter of debate \cite{ns}.
Alternative theories describe the finite D spin glass phase in all
dimensions as a coexistence of two phases related by spin reversal
symmetry \cite{drop}.  The case of generalized spin glasses, or $p$-spin
models, which is interesting in connection with structural glass
physics, also deserves some attention. Mean field theory in fact
predicts the existence of an exponential multiplicity of metastable
states, able to dynamically confine the system in high
free-energy regions. For a long time it has been acknowledged the necessity 
to properly include finite interaction range effects in order
to describe the barrier crossing between states. The understanding of
this phenomenon in Kac models would pose the basis for a
theory of activated processes in finite D glassy systems.

This is the organization of the present contribution: in Section
\ref{sec:def}  the basic models are defined. In section 
\ref{sec:kac} I discuss the behavior of the systems in the Kac
limit and physical implications for systems wit large but finite interaction range.  In Section \ref{sec:ft} we discuss
some numerical results for the 1D Kac spin glass. In Section
\ref{sec:lo} we discuss correlation lengths in $p$-spin models.
Finally, we draw our conclusions.
\section{Spin glasses with Kac-type interactions}
\label{sec:def}
Spin-glass with Kac interactions were first defined in \cite{FZ}. Here
we will refer to the models given in \cite{FTjpa}, where a $p$-spin model
with Kac-like interaction is defined by the Hamiltonian:
\begin{equation}
\label{hP}
H^p(\sigma)=-\sum_{1\le i_1<\cdots< i_{p}\le N}
J_{i_1\cdots i_{p}}
\sigma_{i_1}\cdots\sigma_{i_{p}}
\end{equation}
where $\sigma_i$ ($i=1,...,N$) are Ising or spherical spins in the D
dimensional cubic box $\Lambda=\{1,2,...,L\}^D$ with periodic boundary
conditions (with $N=L^D$) and the couplings $J_{i_1\cdots i_{p}}$ are
Gaussian i.i.d. random variables with zero average and variance
\begin{equation}
  E J^2_{i_1\cdots i_p}=\frac{p!}{R^{Dp}}\sum_{k\in\Lambda}
 \psi(|i_1-k|/R)\cdot...\cdot\psi(|i_2-k|/R)
\label{av}
\end{equation}
where $\psi(x)$ is a range one positive function normalized in a way
that $\int_{R^d} d^dx \; \psi(|x|)=1$. The variable $R$ is the
interaction range: the form (\ref{av}) is a convenient form to let 
only groups of variables within a distance of order
$R$ effectively interact. 

Notice that the interaction is scaled in a way that if $R=L$ one
effectively recover the mean-field model where $E J^2_{i_1\cdots
 i_p}=\frac{p!}{N^{p-1}}$.
\section{The Kac limit}
\label{sec:kac}
\subsection{Free-energy}
In the previous section we have seen that the interaction range $R$ is
taken to be equal to $L$, then one recovers the mean field model. 
The scope of this section is to discuss in an informal way the
physical implication of two theorems concerning the behavior of the
the disordered models (\ref{hP}) in the Kac limit. One
would like then to understand for large $R$, but $R\ll L$. It is clear
that a necessary condition for applicability of mean-field theory to
finite range systems is a smooth crossover from $R\ll L$ to
$R\sim L$.  The first quantity of interest is of course the 
free-energy. Consider the the finite volume -
finite range quenched free-energy, with standard notations: 
\begin{eqnarray}
f_{L,R}(T)=-\frac{T}{N}E\log Z_{L,R}(T;J). 
\end{eqnarray}
One would like to compare the behavior of the infinite volume
free-energy for finite $R$
\begin{eqnarray}
\lim_{R\to\infty}\lim_{L\to\infty}
f_{L,R}(T)
\end{eqnarray}
to the one of the mean field model, for which
\begin{eqnarray}
f^{MF}(T)=\lim_{L\to\infty}
f_{L,L}(T). 
\end{eqnarray}
Let us consider the behavior
of the infinite volume free-energy
$f_{R}(\beta)=\lim_{L\to\infty}f_{L,R}(\beta)$. This is an existing
function, self-averaging with probability one with respect to the
realization of the random couplings. A theorem first proved
in \cite{FTprl}, insures that for all temperatures in the Kac limit
$R\to\infty$ the Mean-Field free-energy is recovered.  This is a
continuity result: the numerical value of the free-energy for large
$R$ is close to the one of the MF function. One may notice that in the
context of spin glass models there are no non-convexities of the
free-energy as a function of the temperature and it does not arise the
need of Maxwell construction.\footnote{Non convex free-energies appear in the
context of $p>2$ models when considering constrained systems
\cite{KPV,FPpot}. Unfortunately, although
there is no particular reason to doubt of the validity of the Maxwell
construction, a formal prove is still to be provided.}

Without entering into the details of the prove let us mention that
this is based on  interpolation inequalities \cite{GT} between the Kac
models and their Mean Field correspondent.
\subsection{Local Spin Glass order}
Of course, as already emphasized by Kac Ulembeck and Hammer \cite{kuh}
in the non disordered context, the continuity of the free-energy has
no implication for the phase diagram at finite $R$, any analytic
function (as e.g. the free-energy as a function of the temperature for
large $R$ in D=1) can be well approximated by a non-analytic one as
provided by the MF limit. However the result suggests that at the
local level, on scales that diverge with the range of interaction, the
physics of finite range models should be well described by mean-field
theory.  Indeed, this is the conclusion of the study of local order
parameter in the Kac limit \cite{FTjpa}.

The spin-glass order parameter, capable to describe ergodicity braking
in the mean-field models is the probability distribution of the
overlap between identical copies of the system \cite{MPV}.  In the context of
finite dimensional, extended systems it is natural to study the
behavior of local overlaps.  One can then define window overlaps on a
box $B_\ell$ of scale $\ell\in (0,L]$ centered around an arbitrary
point: given two spin configurations $\sigma$ and $\tau$,
\begin{equation}
q_\ell (\sigma,\tau )=\frac{1}{|B_\ell |} \sum_{i\in B_\ell} \sigma_i \tau_i.
\end{equation}
The corresponding probability distribution induced by thermal and
quenched disorder is: 
\begin{equation}
  P_{L,\ell} (q)= E\left[ \frac {1}{Z^2}\sum_{\sigma,\tau} 
    e^{-\beta (H(\sigma)+H(\tau) )}  \delta\left( 
q_\ell (\sigma,\tau )-q\right )
\right].
\end{equation}
One would like to understand how the behavior of the order parameter
depends on $\ell$. Mean-Field long-range order would correspond to
MF-like probability distribution functions (PDF) at the largest scale
$L$. A fundamental question about the nature of finite D spin glasses
is whether this long-range order is possible. More modestly, one can
investigate the possibility of local MF order and define a (possibly
infinite) correlation length $\xi$ marking the cross-over from a
non-trivial MF-like behavior of the box overlap to a trivial one.

A second result on the Kac limit concerns then box overlaps on the
scale $\ell=R$ of the interaction \cite{FTjpa}. The overlap can be
analyzed through linear response theory \cite{FMPP}, considering 
a generalized model whose Hamiltonian is a sum of the original one
plus small contributions of all $p$
\begin{eqnarray}
H=H^p+\sum_r C_r H^r
\end{eqnarray}
with $C_r\ll 1$ and where each of the $H^r$ is of the Kac kind for same
range of interaction.  For any $L$ and $R$, the derivative of the
free-energy with respect to the couplings $C_r$ generate the moments
of the overlap distribution:
\begin{equation}
\frac{\partial }{\partial C_r} f_{L,R}(T)=-\beta 
\left(1-\int P_{L,R}(q)q^r\right).
\end{equation}
It was proved in \cite{FTjpa} that for almost all choices of the
parameters $C_r$ in probabilistic sense, the function $P_{L,R}(q)$
tends to the corresponding mean-field function in the Kac limit.  The
main implication of this result is that for large but finite $R$ and
infinite $L$ at least on a local level on scales of order $R$ mean
field order holds. This puts a lower bounds to the overlap correlation length
\begin{eqnarray}
\xi\geq R.
\end{eqnarray}
\subsection{Coarse graining and replica field theory}
In order to understand genuinely finite dimensional systems one needs
to go beyond the Kac limit and study possibly large, but finite values
of $R$.  Unfortunately, in this case mathematically rigorous analysis
becomes prohibitively complicated. One has to resort then to the
available non rigorous techniques of theoretical physics. In
\cite{FTjstat} the problem was addressed with the replica method.  It
was shown there that it is possible to coarse grain the space on
scales $\ell$ $1\ll \ell\ll R\ll L$ such that the $n$-times replicated
partition function can be written as a functional integral of the
exponential of an action for local overlap $n\times n$ matrices
$Q_{ab}(x)$ on the coarse graining scales.
\begin{equation}
E(Z^n)=\int {\cal D}Q_{ab}(x)\exp \left(-R^D \int d^D x \;S[Q,x] \right). 
\label{zn}
\end{equation} 
The coarse grained action $S$ has been computed explicitly in
\cite{FTjstat} and turns out to be independent of
$R$. Abstracting from the functional form of $S$ an interesting
aspect of formula (\ref{zn}) is the appearance of ``the volume of
interaction'' $R^D$ in front of the action. This suggests to treat
finite dimensional effects through asymptotic expansions in $1/R$
based on instanton techniques where one seeks for non spatially
homogeneous saddle points $S$.
\section{1D Spin glasses.}
\label{sec:ft}
We consider in this section the spin glass model $p=2$ in dimension
D=1.  In this case for any finite $R$ there is no phase transition:
the equilibrium phase is paramagnetic and correspondingly the overlap
correlation length $\xi_R$ stays finite at all temperatures. Studying
the window overlap one should then observe at low temperature a
cross-over from a spin-glass regime for to a paramagnetic one as the
window size is increased. An interesting question concerns the
behavior of the overlap correlation length as a function of $R$ below
the mean-field critical temperature, specifically is its growth linear
in $R$ as the analysis of the Kac limit suggests, or is it instead
super-linear? This would appear as the simplest theoretical question to
go beyond the Kac limit. Yet it has not been answered analytically in
the low temperature phase so far. I would like to present here some
results in this sense obtained through numerical simulations in
\cite{FPepl}.

The $p=2$ model (\ref{hP}), where each spin interacts in 1D with order
$R$ neighbors with a strength of order $1/\sqrt R$ is not very
suitable for simulational purposes. A better choice of a model with
a similar low temperature behavior
consists in letting each spin
$\sigma_i$ ($i=1,...,L$) to interact with a small number of other
spins $j$ randomly chosen within the neighborhood $|i-j|\leq R$ which has 
been analyzed in the Kac limit in \cite{GTdil,FTdil}.

\begin{figure}
\resizebox{0.75\columnwidth}{!}{%
  \includegraphics{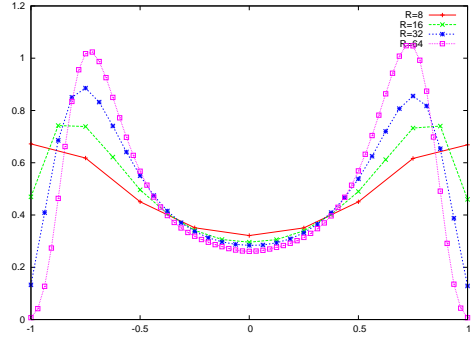}
}
\caption{Distribution of the window overlap on scales $ R$ for
  $T=0.714$ and $R=8,16,32,64$. Increasing $R$ the overlap PDF
  approaches the mean-field distribution.}
\label{sp1}       
\end{figure}

\begin{figure}
\resizebox{0.75\columnwidth}{!}{%
  \includegraphics{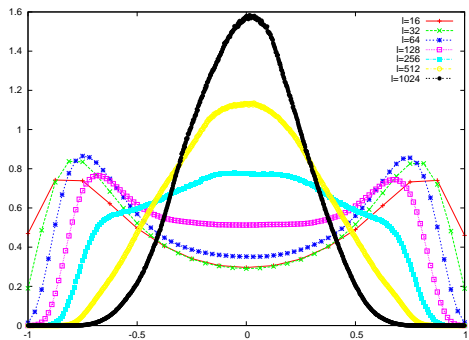}
}
\caption{Distribution of the window overlap on 
scales $\ell $ for $T=0.714$ and $R=16$,  
$\ell=4,8,16,32,64,128,256,512,1024$. It is apparent a cross-over from 
mean-field like to paramagnetic behavior.}
\label{sp2}       
\end{figure}

Figure \ref{sp1} and \ref{sp2} are an illustration of how the analysis
of the previous section applies in a 1D spin glass (the details of the
simulations are given in \cite{FPepl}). In figure \ref{sp1} the window
overlap probability distribution on the scale $R$ is plotted for a
temperature below the mean-field $T_c$. One can see that increasing
$R$ the shape of the curves tend to the characteristic form of mean
field spin glasses, with two symmetric delta peaks at values $\pm
q_{EA}$ and a continuous part for $|q|< q_{EA}$. Figure \ref{sp2} on
the other hand shows the window overlap distribution for the same
temperature, fixed $R=16$ as a function of the window size $\ell$. The
curves, that for small $\ell$ are reminiscent of the mean-field shape,
tend to a Gaussian shape for larger and larger $\ell$.  Quantifying
numerically the growth of the overlap correlation length in the low
temperature phase is a difficult task even in 1D. In \cite{FPepl} a
theoretical and numerical estimate was given for $T=T_c$ where it is
found that
\begin{eqnarray}
\xi\sim R^{6/5}.
\end{eqnarray}
The value of the exponent is larger then the bound implied by the
analysis of the Kac limit. The theoretical derivation \cite{FPepl},
based on dimensional analysis, relies on replica theory, which
predicts that the spin glass transition is governed by a cubic theory
for $T>T_c$ \cite{CL}.

\section{$p$-spin models} 
\label{sec:lo}

The last years have seen a great interest of the statistical physics
community in the structural glass transition. With at the basis a deep
phenomenological analogy, disordered models of the family of the
$p$-spin have been proposed as prototypical systems to understand the
structural glass transition. Indeed they are capable to describe in a
unified framework Mode-Coupling dynamics, the Kauzmann transition as
well as the growth of dynamical correlations on approaching the glass
transition \cite{KTW}.

The physics of $p$-spin mean-field models is dominated by metastable
states. The dynamical Mode-Coupling-like transition at $T_c$ and the
static Kauzmann like transition at the lower transition $T_K$ reflect
the structure of these states. It is a fundamental problem to
understand how the picture extends when non-mean-field effects are
taken into account. In particular it is necessary to understand how
the barriers, which are $O(volume)$ in mean field get modified in
finite $D$, and what are the mechanisms responsible for barrier
crossing. The introduction of Kac models offer the possibility to
study finite dimensional effect in an asymptotic expansion in $R$
around mean field. The barriers in this case, remain finite in the
thermodynamic limit, but diverge as $O(R^D)$ for large $R$
\cite{SFjstat}. Thanks to this fact, for large $R$ one can test in a
controlled theoretical setting various phenomenological ideas that
have been proposed to cure mean-field pathologies in finite dimension.
In particular, it has been possible to study the behavior of two
correlation lengths that have been proposed to be important for the
glass transition \cite{FraMo}.

The study of metastability and relaxation time in glassy systems for
large $R$ has been related in \cite{SFepl} to a free-energy difference
between two systems subject to physical constraints, in analogy with
the classical analysis of metastability of \cite{lp2}.

The idea of \cite{SFepl} is that dynamics proceeds as a passage from
metastable state to metastable state and almost all relevant low
temperature configurations can be taken as representative of a
metastable state. Almost all the other configurations inside the
metastable state, will have in all point of space a local overlap
with the reference above a threshold value. 

Barrier states, allowing the relaxation are at the border of a
metastable state, consist then in configurations where the overlap
with the reference is lower then threshold at least in one point in
space.  These barrier states can be computed considering considering a
system in finite geometry (e.g. a sphere) constrained to have a high
overlap with a reference on the boundary. 

This analysis thus connects barrier to the
``point-to-set'' (PS) correlation functions recently proposed to
identify correlation lengths in glassy systems in alternative to
dynamical determinations \cite{BBps}.

The PS functions measure the correlation in a region of size $\ell$
with a reference configuration fixed as boundary condition outside the
region. The analysis of the PS correlations in p-spin in the Kac limit
in \cite{FraMo} allowed to identify two different relevant lengths
which, for temperatures above $T_K$, scale linearly with the
interaction range $R$.  A first length $\xi_{MC}$, identified as the
dominant length appearing in the dynamical four point function
\cite{length,BB} above $T_c$ and diverging as $R |T-T_c|^{1/4}$,
represents the typical size of regions that can relax in time
independent of $R$ without needing activation.  A second length
$\xi_{Mos}$, diverging at $T_K$ as $R |T-T_K|^{-1}$ represents the
minimal size of regions that can relax through activation, on time
scales exponentially divergent with $R^d$. Below $T_K$ it can be
argued that the dominant relaxation mechanism operate on that scale. 

The two lengths reflect then 
different relaxation mechanisms, with characteristic times displaying
critical mode-coupling and activated scaling respectively:
\begin{eqnarray}
  \tau_{MC}&\sim& \xi_{MC}^z \sim |T-T_c|^{z\nu}
  \\
  \tau_{Mos}&\sim &\exp\left( R^d (\xi_{Mos}/R)^{\psi}C\right)
 \sim \exp\left( 
R^d |T-T_K|^{-\psi} 
    C' \right)
\nonumber
\end{eqnarray}
where $z$ is the dynamical exponent, and the value of the exponent
$\psi = D-1$ in mean field. We see that $\tau_{Mos}$ has a form
similar to the the Vogel-Fulcher form $\tau\sim \exp(C/|T-T_K|)$
expected in 3 dimensions but with a different exponent \cite{KTW}. 

Finally I would like to comment that while the Kac limit provides a
limiting case where analytic progress is possible, the relevance of
the large $R$ analysis for systems with short range interaction is far
from being obvious. This problem has been addressed in the context of
a simple 1D model in \cite{FPR-T}, where it was found that it is only
comparing the behavior of systems with different values of $R$ that
the large $R$ scenario could be verified.

\section{Conclusions}

This paper presents a rapid excursus of my work, in various
collaborations, on disordered models with Kac interactions. 
The main results presented are 
\begin{itemize}
\item The rigorous analysis of the Kac limit. This shows that Mean
  Field theory provides an appropriate description of local properties
  of disordered systems. This is a necessary condition for Long Range
  order to hold, but it does not imply it.
\item Numerical simulations and theoretical arguments show that even
  in $D=1$ where for finite $R$ the system is paramagnetic at all
  temperatures, the correlation length below and at mean field
  critical temperature $T_c$ grows with $R$ faster than the linear
  bound implied by the analysis of the Kac limit.
\item Kac models can be investigate to test in a controlled setting
  phenomenological extension of mean-field theory to deal with
  metastable states in disordered finite dimensional systems. In that
  framework, two relevant lengths with different temperature
  dependence emerge, that describe the size of cooperatively
  rearranging regions in different domains.
\end{itemize}

{\bf Acknowledgments}

Many of the results presented in this paper have been obtained in
different collaborations with A. Montanari, G. Parisi, F. Ricci-Tersenghi,
F.L. Toninelli, whom I warmly thank.

\end{document}